\documentclass[aps,showpacs]{revtex4}
\usepackage{graphicx}

\def\pdif#1#2{\frac{\partial #1}{\partial #2}}

\def\ppdif#1#2{\frac{\partial^2 #1}{\partial #2^2}}

\begin{document}
\title{Quasiequilibrium sequences of synchronously rotating binary neutron 
stars with constant rest masses in general relativity \\
--- Another approach without using the conformally flat condition ---}
\author{Fumihiko Usui}
\email{usui@provence.c.u-tokyo.ac.jp}
\author{Yoshiharu Eriguchi}
\email{eriguchi@valis.c.u-tokyo.ac.jp}
\affiliation{Department of Earth Science and Astronomy,
Graduate School of Arts and Sciences,
University of Tokyo, Komaba 3-8-1, Meguro, Tokyo 153-8902, Japan}
\date{\today}

\begin{abstract}
We have computed quasiequilibrium sequences of synchronously
rotating compact binary star systems with constant rest masses.
This computation has been carried out by using the numerical scheme 
which is different from the scheme based on the conformally flat assumption 
about the space. 

Stars are assumed to be polytropes with polytropic indices of $N=0.5$, 
$N=1.0$, and $N=1.5$. Since we have computed binary star sequences with
a constant rest mass, they provide approximate evolutionary tracks of
binary star systems.  For relatively stiff equations of state ($N < 1.0$), 
there appear turning points along the quasiequilibrium sequences plotted in 
the angular momentum --- angular velocity plane. Consequently secular
instability against exciting internal motion sets in at those points.
Qualitatively, these results agree with those of Baumgarte {\it et al.} who 
employed the conformally flat condition. 

We further discuss the effect of different equations of state and 
different strength of gravity by introducing two kinds of dimensionless 
quantities which represent the angular momentum and the angular velocity. 
Strength of gravity is renormalized in these quantities so that the quantities 
are transformed to values around unity. Therefore we can clearly see
relations among quasiequilibrium sequences for a wide variety of 
strength of gravity and for different compressibility.
\end{abstract}


\maketitle

\widetext


\section{Introduction}

Binary neutron stars are very interesting objects. From the observational 
point of view, we will have a chance to get new eyes for the Universe by 
detecting gravitational waves in the first or second decade of this century. 
It is highly possible that the first signal may be that from compact binary 
stars, such as binary neutron stars, a black hole --- neutron star binary 
system, or binary black holes. On the other hand, theoretically, our 
understanding of evolution of compact binary stars is far from complete 
because it is considerably difficult to treat the ``2-body'' problem from a 
state with a wide separation to a merging stage consistently in the framework 
of general relativity.

However, recent investigations have found a new approach to this problem.
Since the time scale of the orbital change due to gravitational wave emission 
is rather long compared with the orbital period except for in the final few 
milliseconds of the coalescing stage, we can neglect gravitational wave 
emission for most stages of evolution. In other words, we can treat the 
system in ``quasiequilibrium'' (see e.g., \cite{WMM96,BCSST98a,BGM99}).

Following this idea, several groups have obtained quasiequilibrium sequences 
of binary neutron 
stars~\cite{WMM96,BCSST98a,BGM99,UE00,UUE00}. Most of them adopted the 
assumption that the spatial part of the metric is conformally flat (the 
conformally flat condition: hereafter, CFC)~\cite{WMM96,BCSST98a,BGM99,UE00}.
For axisymmetric rotating polytropes results of the scheme with the CFC were 
compared with those obtained by the numerically exact code and found to be
reasonably accurate~\cite{CST96}. However, since there are no exact numerical 
solutions for binary configurations, one could not know the accuracy of the
results obtained by the scheme with the CFC. 

Therefore, it is desirable to develop different schemes from that with the CFC 
and to compare results of different schemes for nonaxisymmetric 
configurations. As one of those alternatives, in the previous paper,
we developed a new numerical scheme to obtain quasiequilibrium structures 
of nonaxisymmetric compact stars as well as the space time around those stars 
in general relativity and obtained quasiequilibrium sequences of synchronously 
rotating binary polytropes~\cite{UUE00}. In that scheme, the Einstein 
equations are solved directly without assuming the CFC. The obtained results, 
however, could not be compared with those of \cite{BCSST98a} because 
different polytropic relations were used. 

In this paper, we have used the same polytropic equation as that used 
in~\cite{BCSST98a} and computed quasiequilibrium sequences of synchronously
rotating polytropes with a constant rest mass. Therefore, we can directly 
compare our results with those with the CFC.  In actual computations, we have 
improved our numerical scheme and succeeded in making our scheme more robust.

\section{Assumptions and basic equations}
\subsection{Assumptions}

As mentioned in Introduction, we can neglect the effect of gravitational wave
emission for almost all stages except for the last few milliseconds of the 
coalescence of binary systems and choose a rotating frame in which 
the system is in a stationary state.

In this paper, we use the units of $c=G=1$ except for in the final section
and we will make the following assumptions (see, Usui, Ury\=u, and 
Eriguchi~\cite{UUE00} for a more detail):

\begin{description}
\item [(1)] We deal with a binary system which consists of two stars of 
equal masses in a circular orbit.
\item [(2)] The binary star system is in a stationary state in the rotating 
frame with the orbital angular velocity $\Omega$.
\item [(3)] Axes of spins of two stars and that of the orbital motion are 
parallel to one another. A schematic figure of the system is shown in 
Fig.~\ref{coordinate system}.
\item [(4)] Spins of two stars are synchronized to the orbital motion.
Each star is rigidly rotating with the angular velocity $\Omega$ if seen 
from a distant place. 
\item [(5)] The matter of the star is a perfect fluid and
the equation of state for the matter is assumed to be polytropic:
\begin{eqnarray}
\label{polytrope}
p &=& K \rho ^{1+1/N} \ ,\\
\varepsilon &=& \rho + Np \ ,
\end{eqnarray}
where $p$, $\rho$, $\varepsilon$, $N$, and $K$ are the pressure, the rest mass 
density, the energy density, the polytropic index, and the polytropic 
constant, respectively.
\end{description}

\subsection{Metric and basic equations}
\label{formulation}

In addition to the assumptions mentioned above, we further assume the 
following form for the metric in the spherical coordinates 
$(r, \theta, \varphi)$ :
\begin{equation}
\label{metric}
ds^2  = - e^{2\nu}dt^2
     	 + r^2\sin^2\theta e^{2\beta}(d\varphi -\omega dt)^2
         + e^{2\alpha}dr^2  
	 + r^2 e^{2\alpha}d\theta^2 \ , 
\end{equation}
where $\nu$, $\beta$, $\omega$, and $\alpha$ are the metric
potentials which are functions of $r$, $\theta$, and $\varphi$.  
Although this form of the metric becomes exact for stationary axisymmetric 
configurations, it is not exact for nonaxisymmetric configurations.
Therefore, it must be extended to a more general form in the future.

The Einstein equations  with appropriate boundary conditions can be written 
down for the metric form Eq.~(\ref{metric}), and are transformed to integral 
representations by using the Green function for the Laplacian in the flat 
space (see~\cite{UUE00}).
Here we solve four equations for the metric potentials, $\nu$, $\beta$, 
$\omega$, and $\alpha$, and the rest of the Einstein equations are not used.
In the concrete, we do not consider ${(t)(r)}$, ${(t)(\theta)}$, 
${(r)(\varphi)}$, and ${(\theta)(\varphi)}$ components 
of the Einstein equations (see~\cite{UUE00}). 
We can check how these equations are satisfied or violated after obtaining
the solutions to the rest of the Einstein equations. In actual computations, 
even for highly relativistic configurations, these equations are satisfied 
quite well, to within $0.5\%$, if we estimate the relative errors by
the following quantity:
\begin{eqnarray}
\frac{R_{(\mu)(\nu)} - 8 \pi \left(T_{(\mu)(\nu)} 
- 1/2 g_{(\mu)(\nu)} T \right) }
{\sum{\Bigl|\mbox{each term of }R_{{(\mu)}{(\nu)}}\Bigr|}} \ ,
\end{eqnarray}
where $R_{(\mu)(\nu)}, T_{(\mu)(\nu)}$ and $T$ are the Ricci tensor, the energy
momentum tensor and the trace of the energy momentum tensor, respectively.
It must be noted that for nonaxisymmetric quasiequilibrium configurations
it is {\it impossible} to satisfy all components of the Einstein equations 
because there cannot exist ``exact'' quasiequilibrium configurations
as far as the asymptotic flat condition is assumed.

\section{Method of calculations}

\subsection{Solving scheme}

The method of calculation is almost the same as that used in the previous
paper~\cite{UUE00} except for several improvements. We will briefly summarize
the scheme and explain the improved changes in some detail.

\begin{figure}[htbc]
\includegraphics[width=0.5\textwidth]{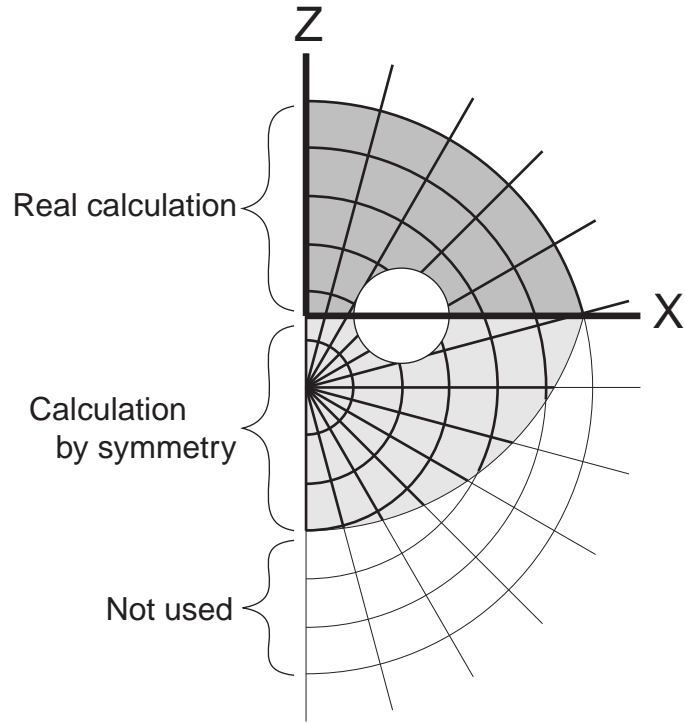}
\caption{Shifted coordinate system. A circle located on the
$x$-axis denotes a component star of the binary.
\label{shifted coordinate system}}
\end{figure}

We solve the hydrostatic equation and the metric potentials 
iteratively by adopting the HSCF (Hachisu's Self-Consistent Field) 
method~\cite{UUE00,H86,KEH89}. In particular, as mentioned before, 
the Einstein equations for the metric are transformed into the integral 
equations by constructing the Laplacian in the flat space. In this procedure
we have to add the following derivative to both sides of the equation:
\begin{equation}
\frac{1}{r^2 \sin^2\theta} \ppdif{g_{{\mu}{\nu}}}{\varphi} \ .
\end{equation}
This term may cause the iteration diverge because, once a certain
amount of numerical errors is introduced by chance, this error may not 
decrease but increase iteration by iteration near the coordinate center. 
To avoid this problem, we have shifted the coordinate center downwards 
along the $z$-axis as Fig.~\ref{shifted coordinate system}. 
Numerical computations are carried out only in the upper half of the whole 
space, i.e., above the equatorial plane of the binary star systems, and 
quantities in the lower half are computed by making use of the symmetry of 
the physical quantities about the equatorial plane. Consequently, it is noted 
that 25\% of the whole grid points is used in the actual calculations.
As shown later, this scheme works nicely.

We have used $(r \times \theta \times \varphi) = (200 \times 119 \times 81)$ 
grid points for ($0 \leq \tilde{r} \leq 2.0$, $0 \leq \theta \leq \pi$, 
$0 \leq \varphi \leq \pi/2$), where $\tilde{r} = r / r_{B}$, and $r_B$ means 
the distance from the orbital axis to point B 
(see Fig.~\ref{coordinate system}).

\subsection{Physical quantities}
\label{Physical quantities}
Since there exist no exact equilibrium or stationary states for real compact
binary stars with a flat space at infinity, it is difficult to define exactly 
the conserved quantities for nonaxisymmetric configurations in general 
relativity. However, for an asymptotically flat spacetime, we can define the 
{\it approximate} total angular momentum, $J_{\rm tot}$, and 
the {\it approximate} total gravitational mass, $M_{\rm tot}$, as:
\begin{eqnarray}
\label{angular momentum}
J_{\rm tot} &=& \int r\sin\theta e^{2\alpha+2\beta}\frac{(\varepsilon+p)v}{1-v^2} r^2\sin\theta dr d\theta d\varphi\nonumber\\
& &{}
+\frac{1}{4\pi}\int e^{2\alpha} \frac{v}{r \sin \theta} 
\left[\ppdif{\alpha}{\varphi}
+\pdif{\alpha}{\varphi}\left(\pdif{\alpha}{\varphi}-\pdif{\beta}{\varphi}-\pdif{\nu}{\varphi}\right)\right]r^2\sin\theta dr d\theta d\varphi \ ,\\
M_{\rm tot} &=& \int e^{2\alpha+\beta+\nu}\left\{(\varepsilon +p)\frac{1+v^2}{1-v^2}+2p\right\} r^2\sin\theta dr d\theta d\varphi\nonumber\\
& &{}
+\int 2r\sin\theta e^{2\alpha+2\beta}(\varepsilon +p)\frac{v\omega}{1-v^2}r^2\sin\theta dr d\theta d\varphi \ ,
\end{eqnarray}
where $v = r (\Omega - \omega) \sin \theta e^{\beta - \nu} $ is the velocity 
of the matter.
These expressions can be obtained by making use of the behaviors of the metric
functions at flat infinity as follows (see~\cite{UUE00} for a more detail): 
\begin{equation}
\omega \sim 2J_{\rm tot}/r^3\ ,\
\partial\omega/\partial r\sim - 6J_{\rm tot}/r^4\ ,\
\partial{\nu}/\partial{r} \sim {M_{\rm tot}}/{r^2}, \
{\rm at}\  r \sim \infty  \ .
\end{equation}

Notice that these expressions are not general forms and different from the
ADM quantities used in Baumgarte {\it et al.}~\protect\cite{BCSST98a}.
To compare our results with those with the CFC, it would be desirable to 
employ the same definitions of physical quantities as theirs. For conformally
flat situations, the ADM mass can be easily evaluated by using the
matter contribution in addition to the contribution of the extrinsic 
curvatures as seen in Baumgarte {\it et al.}~\protect~\cite{BCSST98a}.
With our choice of the metric (Eq.~(\ref{metric})), however, it is difficult
to use the same definition because we do not use the conformal factor. 
Therefore we follow the scheme used by Bardeen~\cite{B73} and derive
the total mass and the total angular momentum by considering the
asymptotic behavior of the 4-metric.
In the previous paper~\cite{UUE00}, the expression of the angular momentum 
(Eq.(48) in \cite{UUE00}) was incorrect. Numerical values of the angular
momentum in~\cite{UUE00} contain about
$15\%$ errors at maximum
due to that wrong expression. 

On the other hand, for the rest mass, it is natural to define as:
\begin{eqnarray}
M_{\rm 0,tot} &=& \int \rho u^t \sqrt{-g} \cdot r^2\sin\theta dr d\theta d\varphi\nonumber\\
&=& \int e^{2\alpha+\beta} \rho \frac{1}{\sqrt{1-v^2}} 
r^2\sin\theta dr d\theta d\varphi \ .
\end{eqnarray}

Since we use the same polytropic relation as that used by Baumgarte 
{\it et al.}~\cite{BCSST98a},  we can choose the same nondimensional 
quantities as they did by using the polytropic constant $K$ and 
the polytropic index $N$ as:
\begin{eqnarray}
\label{nondimensional M}
M &=& K^{N/2} \bar{M} \ ,\\
M_0 &=& K^{N/2} \bar{M_0} \ ,\\
\label{nondimensional e}
\varepsilon &=& K^{-N} \bar{\varepsilon} \ ,\\
\label{nondimensional omega}
\Omega &=& K^{-N/2} \bar{\Omega} \ ,\\
\label{nondimensional J}
J &=& K^N \bar{J} \ ,
\end{eqnarray}
where $M$, $M_0$  and $J$ are the gravitational mass, the rest mass and 
the angular momentum of 
a component star and quantities with ``bar'' mean nondimensional ones.

\subsection{Model parameters}

\begin{figure}[htbc]
\includegraphics[width=0.5\textwidth]{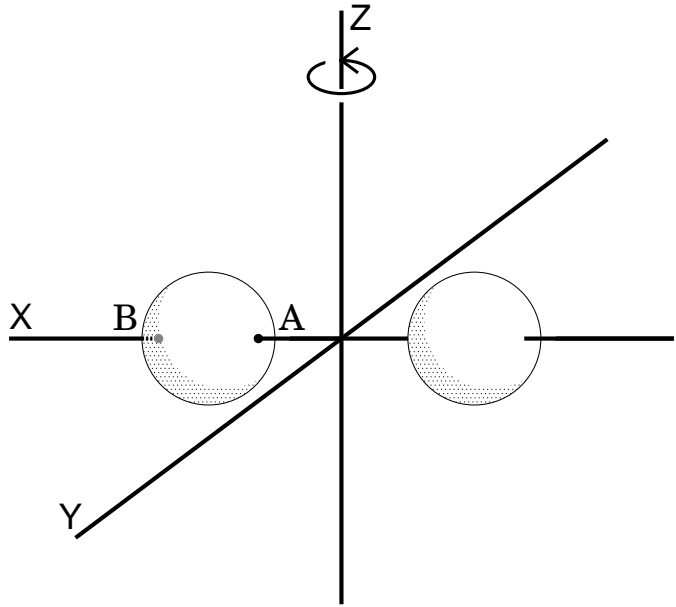}
\caption{Schematic view of the binary system and the coordinates. 
The $x$ axis is set along the line joining the two centers of mass of the two
stars and the origin is set at their midpoint. The $z$ axis is along
the rotation axis. Points A, B are set at the intersections of the surface 
of the star and the $x$ axis. The inner intersection is point A and 
the outer one is point B.\label{coordinate system}}
\end{figure}

For polytropes, the equation of state is fixed by choosing one parameter, 
constant $K$, in addition to the polytropic index $N$.  
After the equation of state is fixed, we need to specify two more parameters
to determine one rotating equilibrium configuration:
one represents the strength of gravity and the other for the rotation.
In our formulation, we choose the following two parameters:
\begin{description}
\item [(1)] The maximum energy density, $\varepsilon_c$.
\item [(2)] The ratio of the shortest distance $r_A$ (distance from the 
orbital axis to point A, see Fig.~\ref{coordinate system}) to the largest 
distance $r_B$ (distance from the orbital axis to point B, 
see Fig.~\ref{coordinate system}), $q$:
\begin{equation}
\label{q define}
q \equiv \frac{r_A}{r_B} \ .
\end{equation}

This quantity $q$ indirectly specifies the rotational state and is related 
to the separation --- distance between the two stars --- as $d = r_B(1+q)$, 
where $d$ is the distance between the geometrical centers of the stars.
\end{description}

Since the quantity $K$ does not appear in the basic equations
due to our normalization, i.e., 
Eqs.~(\ref{nondimensional M})--(\ref{nondimensional J}),  we can treat 
the problem by specifying only three parameters, $\varepsilon_c$ (or 
nondimensional quantity $\bar{\varepsilon}_c$) and $q$, in addition to the 
polytropic index $N$.

\section{Results}
\subsection{Construction of evolutionary sequences of binaries with constant
rest masses}

In order to consider realistic evolutions of compact binary star systems, 
it would be important to find conserved quantities which characterize the
evolution. Although, during evolutions of binary neutron star systems, the
gravitational mass and the angular momentum are lost from the system,
the baryon number of the system should be conserved because we do not 
consider the mass loss or the Roche lobe overflow.  Therefore a sequence 
of a constant rest mass (baryon mass) can be considered to provide 
an evolutionary track.

Even if the rest mass of the star is specified, one cannot follow an 
evolutionary sequence of binary stars.  One needs to know the following two 
things: 1) the change of the equation of state and 2) the change of the 
velocity field of the stars. Concerning the first point, as far as one uses 
the polytropic relations, it is impossible to know the change of the equation 
of state. Thus we adopt the same approximation as Baumgarte {\it et al.} did, 
i.e., the polytropic index $N$ and the constant $K$ are fixed during binary 
evolutions~\cite{BCSST98a}.  Concerning the value of $K$, it may be a 
reasonable assumption because $K$ means the entropy in the Newtonian limit 
and the evolution can be treated as an almost adiabatic process, i.e., the 
timescale of viscous heating is much longer than evolutionary time scale due 
to the back reaction of gravitational wave emission.

As for the second point, it is widely believed that binary neutron star 
systems evolve irrotationally~\cite{K92,BC92}. However, since the purpose
of the present paper is to present an alternative method of solving
binary neutron stars without employing the CFC, we assume that the system 
evolves by keeping synchronous rotation, although its evolution is not
realistic.

Although our final goal in this paper is to obtain evolutionary sequences of 
constant rest masses for the specified polytropic relation, it is not easy to
follow constant rest mass sequences because the rest mass is not
a local quantity but a global quantity which can be computed after the 
structure of the binary star is obtained. Thus, we construct evolutionary 
sequences of constant rest masses by the following procedure:
\begin{description}
\item [(1)] Fix the polytropic index $N$.
\item [(2)] Calculate sequences by changing $\bar{\varepsilon}_c$ for a fixed 
value of $q$. After obtaining the structure, compute global physical 
quantities such as the rest mass, the gravitational mass, and the angular 
momentum.
\item [(3)] Repeat Step~(2) by changing the value of $q$.
\item [(4)] Specify one value of the rest mass and interpolate physical 
quantities with different values of $q$ for the specified rest mass 
by using quantities obtained in Steps~(2)-(3). Thus constant rest mass 
sequences are obtained.
\item [(5)] Change the rest mass and repeat Step~(4) and construct sequences 
for different strength of gravity.
\end{description}

Hereafter, we choose the compactness, $(M/R)_{\infty}$, as the parameter of 
the strength of gravity.  Here $M_{\infty}$ and $R_{\infty}$ are 
the gravitational mass and the Schwarzschild radius of a spherical star
with the same value of the rest mass as that of the sequence which we consider.

\subsection{Numerical results}

\subsubsection{$N=1.0$ sequences}

\begin{figure}[htbc]
\includegraphics[width=0.75\textwidth]{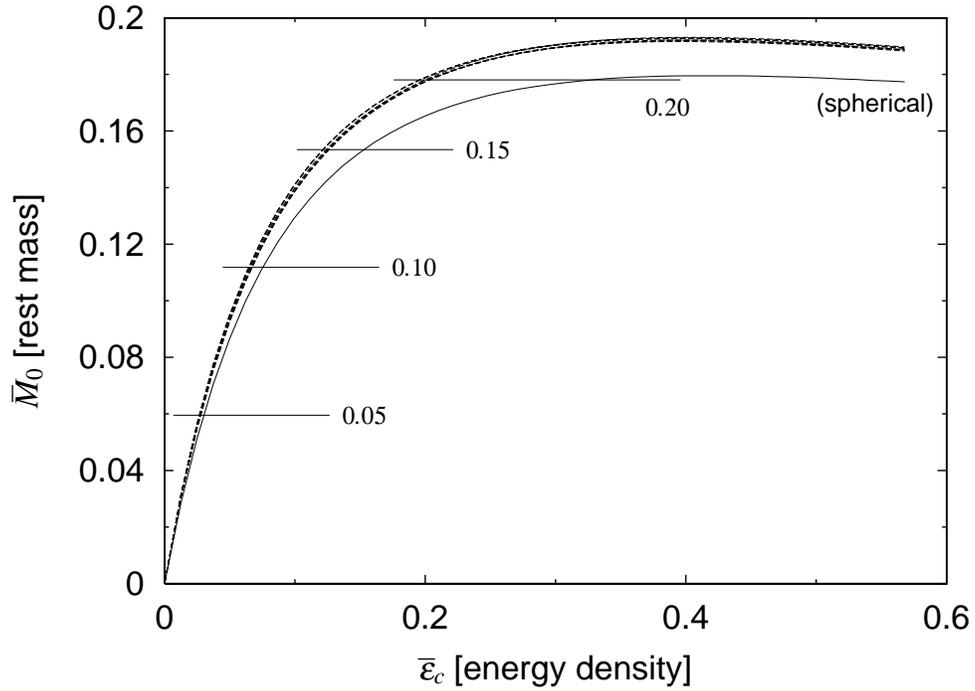}
\caption{Nondimensional rest mass of each star $\bar{M}_0$ is plotted
against nondimensional central energy density $\bar{\varepsilon}_c$ for the 
quasiequilibrium $N=1.0$ polytropic binary star systems with the separation 
$q=0.0$ (contact), $0.05$, $0.1$, $0.15$, $0.2$, $0.25$, and $0.3$.
The curves for binaries with smaller separations locate at upper side, and 
the uppermost curve denotes a sequence with $q=0.0$.
Thin solid curve denotes a sequence of spherical stars with the same polytropic
index. Four horizontal lines labeled by
$(M/R)_{\infty} = 0.05$, $0.10$,  $0.15$, and $0.20$ indicate the constant 
rest mass sequences discussed in Fig.~\protect\ref{J-Omega N=1.0} and
Fig.~\protect\ref{q-dens N=1.0}.\label{dens-mass N=1.0}}
\end{figure}

In Fig.~\ref{dens-mass N=1.0}, the nondimensional rest mass of each star is 
plotted against the nondimensional central energy density. Each curve 
represents a sequence along which the separation $q$ is fixed but the energy 
density  $\bar{\varepsilon}_c$ is varied. Seven curves correspond to the 
sequences of $q=0.00$ (contact), $0.05$, $0.10$, $0.15$, $0.20$, $0.25$, 
and $0.30$.
From this figure, we can see that as the central energy density increases, 
the rest mass increases until a certain critical point. However, as the 
density increases further, the mass begins to decrease because of the 
general relativistic effect.
The curves at the upper side correspond to the binary sequences with 
smaller separations and the uppermost curve denotes the sequence at the 
contact phase.
For larger separations, each star can be considered to approach the 
spherical configuration. 

By using the information in this figure, we can consider evolutionary sequences
with constant rest masses because horizontal lines can be regarded as such
sequences. Thus, we choose four evolutionary sequences of four constant rest 
masses which correspond to four different compactnesses,
i.e., $(M/R)_{\infty} = 0.05$, $0.10$, $0.15$, and $0.20$.

\begin{figure}[htbc]
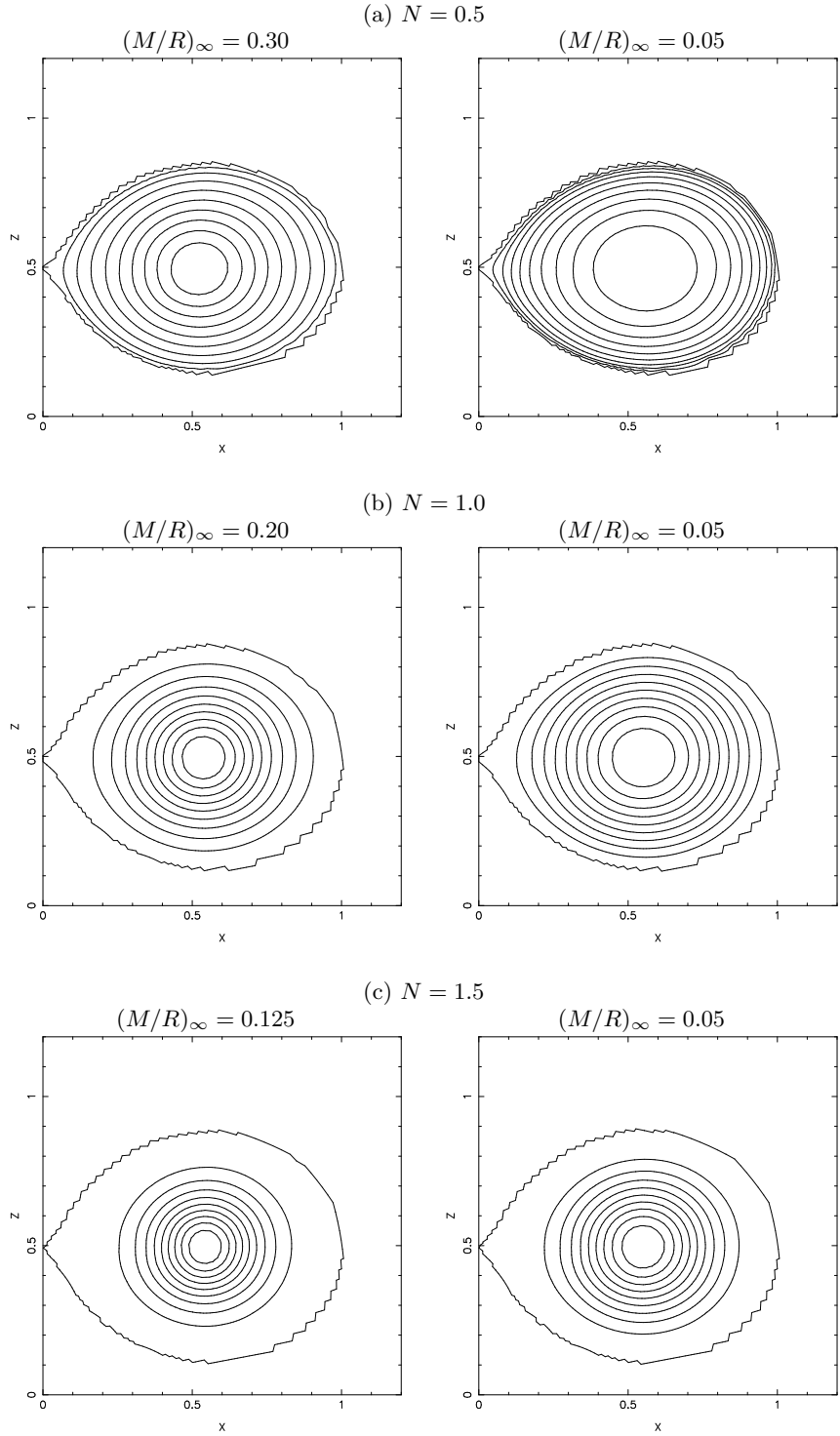

\begin{tabular}{ccccc}
\multicolumn{5}{c}{(a) $N = 0.5$}\\
$(M/R)_{\infty} = 0.30$&&&&$(M/R)_{\infty} = 0.05$\\
\includegraphics[width=0.3\textwidth]{cont.2.005E+15.n0.5.eps}
&&&&
\includegraphics[width=0.3\textwidth]{cont.5.000E+12.n0.5.eps}\\
\\
\multicolumn{5}{c}{(b) $N = 1.0$}\\
$(M/R)_{\infty} = 0.20$&&&&$(M/R)_{\infty} = 0.05$\\
\includegraphics[width=0.3\textwidth]{cont.3.405E+15.n1.0.eps}
&&&&
\includegraphics[width=0.3\textwidth]{cont.5.000E+12.n1.0.eps}\\
\\
\multicolumn{5}{c}{(c) $N = 1.5$}\\
$(M/R)_{\infty} = 0.125$&&&&$(M/R)_{\infty} = 0.05$\\
\includegraphics[width=0.3\textwidth]{cont.3.405E+15.n1.5.eps}
&&&&
\includegraphics[width=0.3\textwidth]{cont.5.000E+12.n1.5.eps}
\end{tabular}
\caption{Contours of the energy density on the $x$-$z$ plane
for models at the contact phase: (a) $N=0.5$, (b) $N=1.0$, and (c) $N=1.5$.
The difference between two adjacent curves is 1/10 of the maximum
energy density.
The units of the distance is $r_B$. For each $N$, contours for two 
values of the strength of gravity are shown: one for less relativistic
models, and the other for highly relativistic models which
correspond to the models with almost the maximum mass of the sequence.
\label{contours}
}
\end{figure}

The energy density contours for selected models at the contact phase
are shown in Fig.~\ref{contours}. In this figure, equidensity contours are 
drawn for several $N$ and $(M/R)_{\infty}$.

In Fig.~\ref{J-Omega N=1.0}, the nondimensional angular velocity is plotted 
against the nondimensional angular momentum.  Four curves correspond to 
sequences with the compactness mentioned above. As the angular momentum 
decreases due to gravitational wave emission, the binary system evolves 
to the upper-left direction along the curve. If the curve has a turning 
point, the system becomes secularly unstable at that point.
As seen from this figure, for $N=1.0$ polytropes, 
it is very difficult to tell whether turning points appear or not,
because turning points, if exist, seem to coincide with the contact states.

In this same figure, the results of Baumgarte 
{\it et al.}~\protect\cite{BCSST98a} are also plotted by using several 
different types of symbols. 
If we assume that the turning points coincide with 
the contact points, we can see almost the same tendency about 
the occurrence of secular instability, although the values are
not the same.
While Baumgarte {\it et al.}~\protect\cite{BCSST98a} used the ADM quantities, 
we have introduced our approximate definitions as mentioned in 
Sec.~\ref{Physical quantities}.
Although these quantities coincide in the Newtonian limit, the 
slight differences appear for highly relativistic, highly nonlinear 
configurations.

\begin{figure}[htbc]
\includegraphics[width=0.75\textwidth]{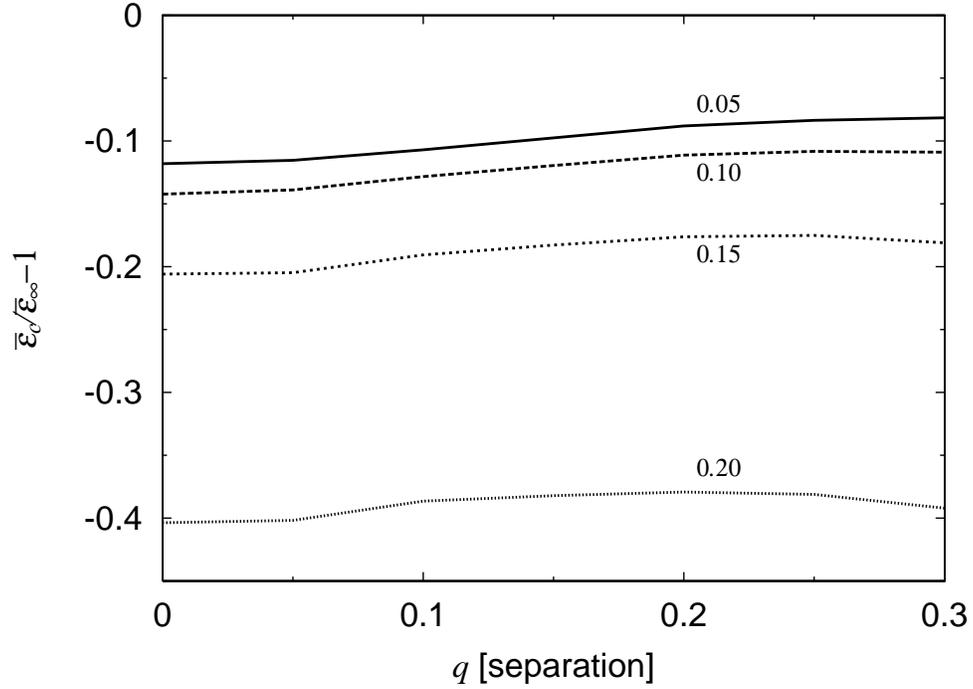}
\caption{Nondimensional central energy density $\bar{\varepsilon}_c$ is 
plotted against the separation $q$ (defined as Eq.~(\protect\ref{q define})) 
for the sequences of constant rest mass polytropic binary systems 
with $N=1.0$. Four curves correspond to sequences with the compactnesses
$(M/R)_{\infty} = 0.05$, $0.10$, $0.15$, and $0.20$.
The quantity $\bar{\varepsilon}_{\infty}$ is the central energy density
of a spherical model.\label{q-dens N=1.0}}
\end{figure}

Fig.~\ref{q-dens N=1.0} shows the nondimensional central energy density 
$\bar{\varepsilon}_c$ --- separation parameter $q$ relation. From this 
figure, we can see that the central energy density decreases as the orbit 
shrinks, i.e., as the evolution proceeds.  The central energy density
seems to decrease as the separation become larger, too. However, this may not
be real because, as mentioned before, a small number of grid points
cover the matter region and values cannot be highly accurate.
Thus, there is no tendency to collapse individually prior to merging,
which was suggested by Wilson {\it et al.}~\cite{WMM96}.

\subsubsection{$N=1.5$ and $N=0.5$ sequences}

\begin{figure}[htbc]
\includegraphics[width=0.75\textwidth]{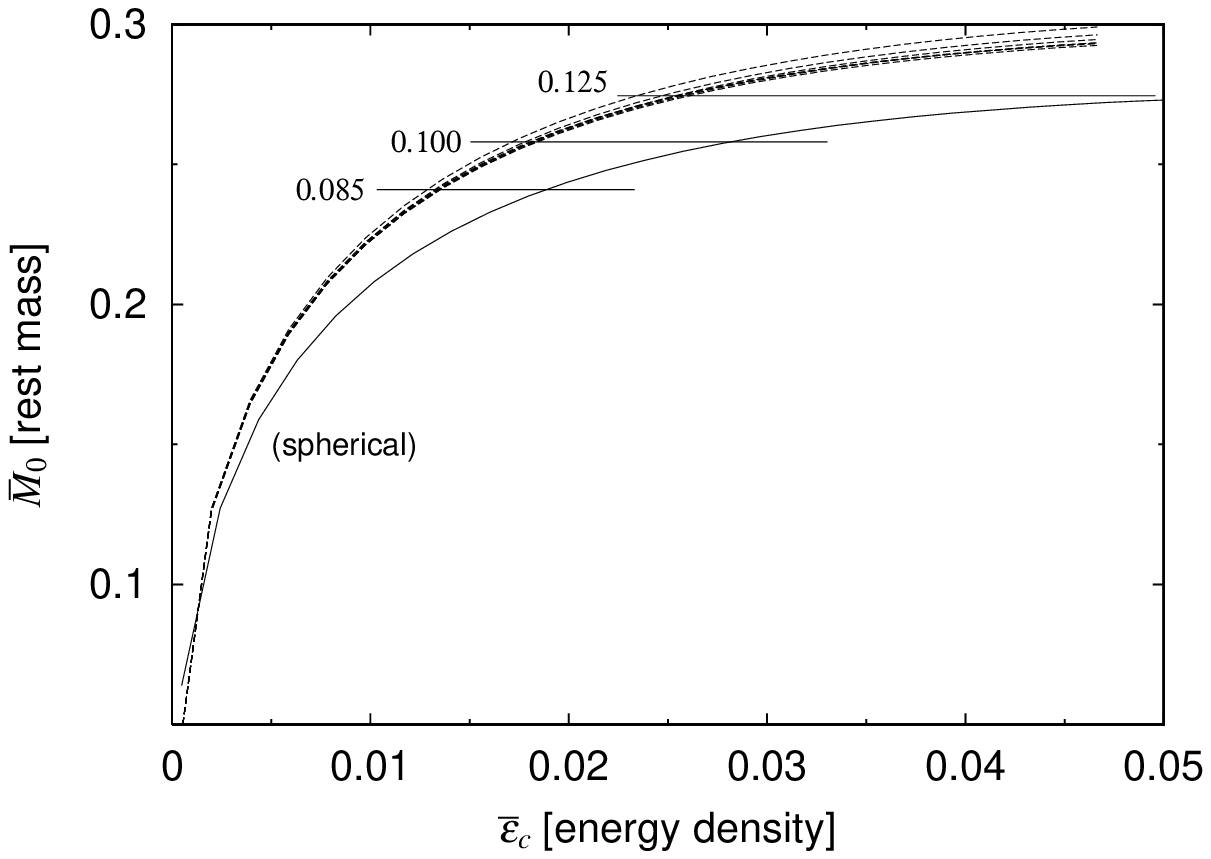}
\caption{Same as Fig.~\protect\ref{dens-mass N=1.0} but for $N=1.5$.\label{dens-mass N=1.5}}
\end{figure}

\begin{figure}[htbc]
\includegraphics[width=0.75\textwidth]{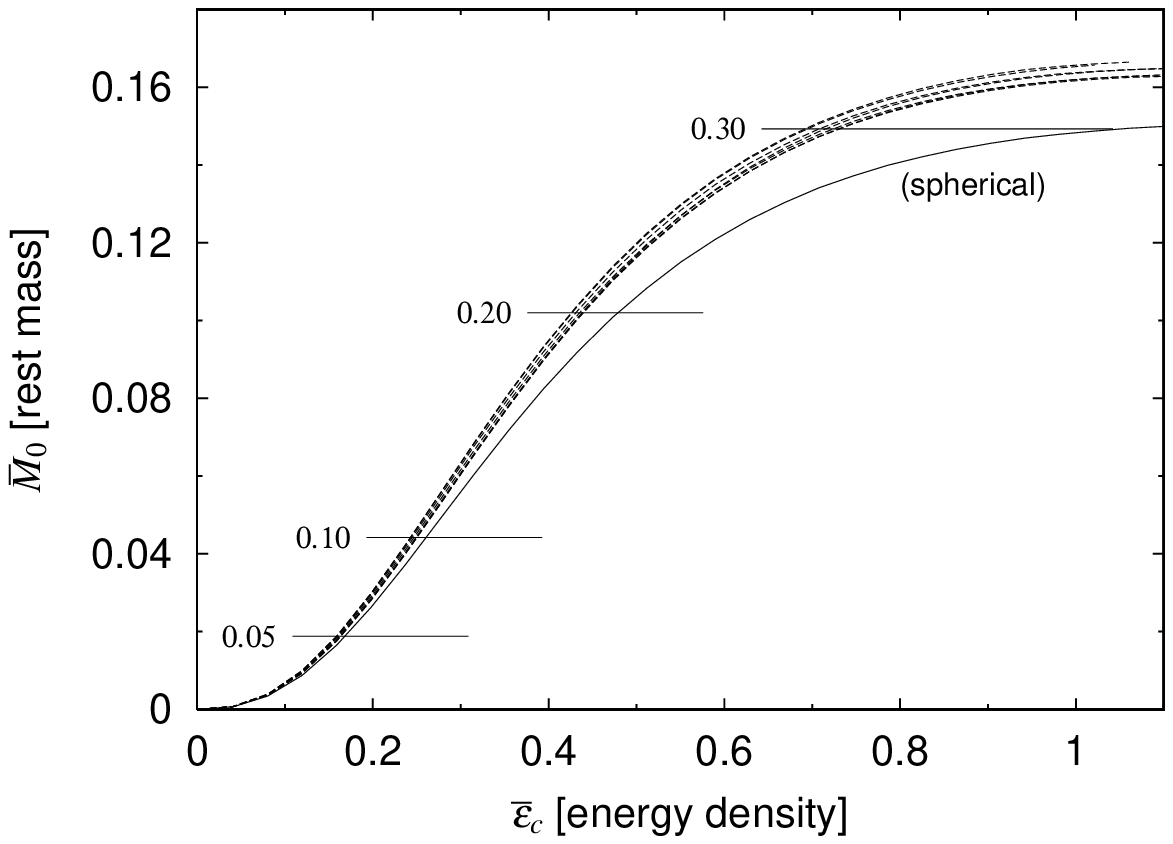}
\caption{Same as Fig.~\protect\ref{dens-mass N=1.0} but for $N=0.5$.
\label{dens-mass N=0.5}}
\end{figure}

In Figs.~\ref{dens-mass N=1.5} and ~\ref{dens-mass N=0.5},
the nondimensional rest mass of each star is plotted against the 
nondimensional central energy density for $N=1.5$ and $N=0.5$ sequences,
respectively. Each curve represents a sequence along which the separation 
$q$ is fixed but the energy density  $\bar{\varepsilon}_c$ is varied. 
The tendency is the same as that for $N=1.0$ sequences.

As for the constant rest mass sequences, we show three selected sequences
with different compactness, i.e., $(M/R)_{\infty} = 0.085$, $0.100$,
and $0.125$ for $N=1.5$ polytropes, and four sequences, i.e., 
$(M/R)_{\infty} = 0.05$, $0.10$, $0.20$ and $0.30$ for $N=0.5$ polytropes.
They are shown in Figs.~\ref{J-Omega N=1.5} and \ref{J-Omega N=0.5}.
As in Fig.~\ref{J-Omega N=1.0}, the nondimensional angular velocity is plotted 
against the nondimensional angular momentum.  
As seen from these figures, for $N > 1.0$ polytropes, the sequences
terminate at the contact phases without encountering the turning
points. It implies that the binary system with a soft equation of state 
evolves to contact states stably. On the other hand, for stiff polytropes, 
$N < 1.0$, the turning points appear before the contact phases so that 
the binary system becomes unstable.

\section{Discussion and Conclusion}

\begin{figure}[htbc]
\includegraphics[width=0.75\textwidth]{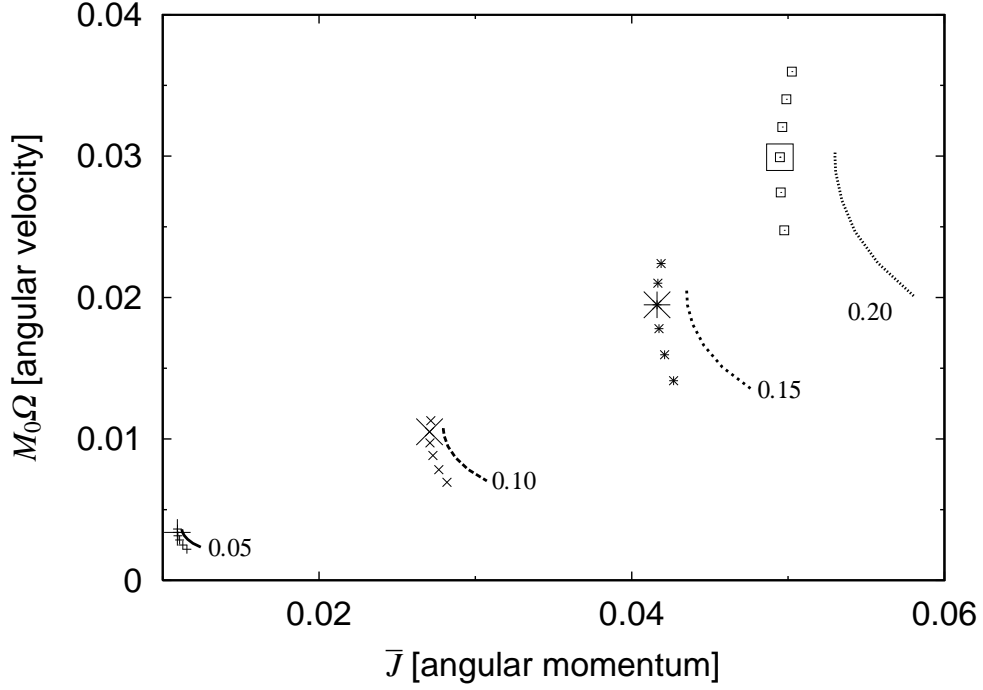}
\caption{Nondimensional angular velocity $M_0 \Omega$ is plotted against 
nondimensional angular momentum $\bar{J}$ for the sequences 
of constant rest mass polytropic binary systems with $N=1.0$.
Four curves correspond to sequences with the compactnesses
$(M/R)_{\infty} = 0.05$, $0.10$, $0.15$, and $0.20$.
Terminal point of each curve at higher values of the angular
velocity corresponds to the configuration at the contact phase.
The results of Baumgarte {\it et al.}~\protect\cite{BCSST98a}
are marked by different symbols and turning points in their results are 
denoted by the same  but larger symbols.
\label{J-Omega N=1.0}}
\end{figure}

\begin{figure}[htbc]
\includegraphics[width=0.75\textwidth]{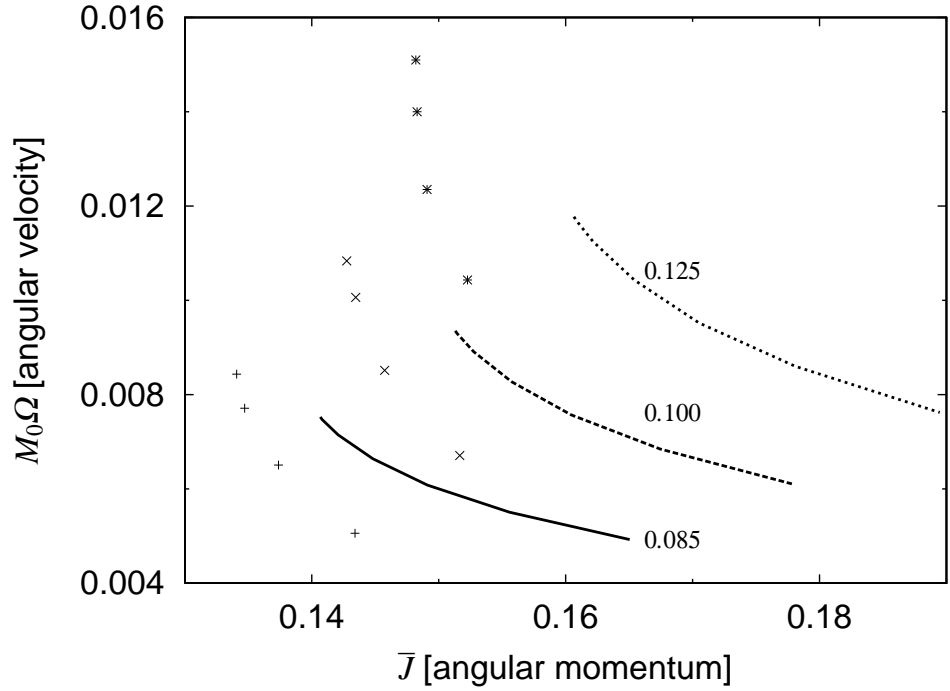}
\caption{Same as Fig.~\protect\ref{J-Omega N=1.0} but for $N=1.5$. 
Three curves correspond to sequences with the compactness
$(M/R)_{\infty} = 0.085$, $0.100$, and $0.125$. Different symbols denote 
the results of Baumgarte {\it et al.}~\protect\cite{BCSST98a}.
\label{J-Omega N=1.5}} 
\end{figure}

In this paper we have solved quasiequilibrium sequences of synchronously 
rotating binary star systems in general relativity without assuming the CFC.
We have constructed the constant rest mass sequences and shown that
for stiff equations of state ($N < 1.0$), evolutionary curves have turning 
points so that synchronous rotation of the system breaks down at that point
and that the internal motion will be excited.
Our results can be compared with those of Baumgarte {\it et al.} who employed 
the CFC~\cite{BCSST98a}.  Quantitatively, there are some differences between
two results as seen from Figs.~\ref{J-Omega N=1.0} and \ref{J-Omega N=1.5}.
These differences may come from different choices of the metric.
However, it should be noted that qualitative features are very similar, 
i.e., the dependency on the polytropic index of the appearance of the 
turning points and so on.  Therefore, although it is hard to give exact
values of the angular velocity and/or the angular momentum at the turning
points from quasiequilibrium approach, the occurrence of the instability
could be correctly predicted. 
Nevertheless, since there are no exact solutions for the binary 
neutron star systems, we should keep in mind that there is a possibility
that both of the two results might not represent the exact solutions.

\begin{figure}[htbc]
\includegraphics[width=0.75\textwidth]{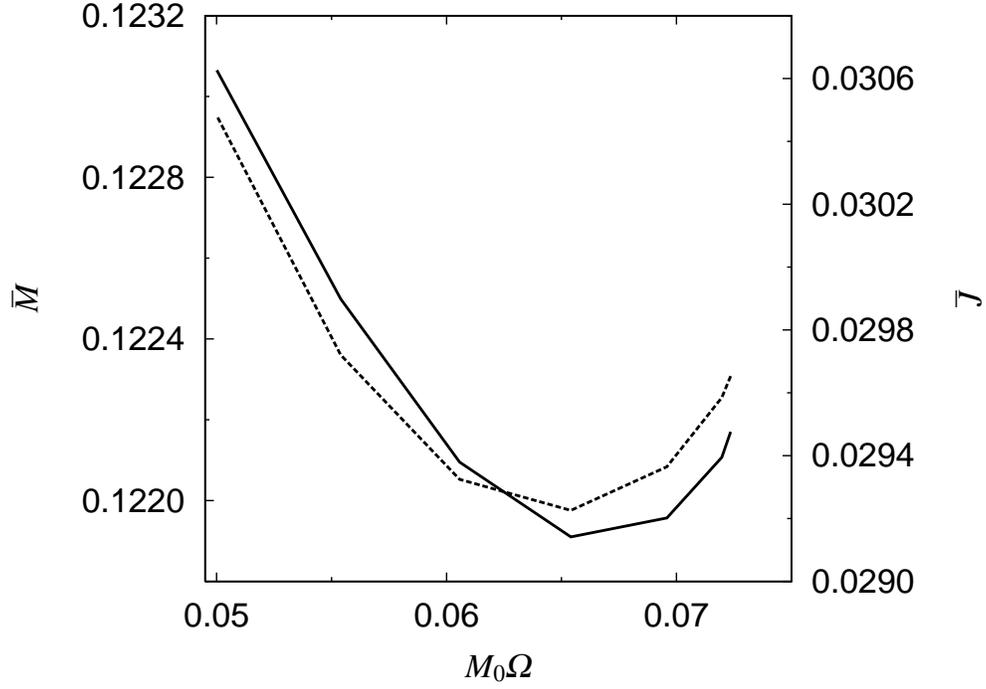}
\caption{Nondimensional gravitational mass $\bar{M}$ (solid curve) and
nondimensional angular momentum $\bar{J}$ (dashed curve)
are plotted against nondimensional angular velocity $M_0 \Omega$ for $N=0.5$,
and $(M/R)_{\infty} = 0.30$.
\label{dj-dm N=0.5} 
}
\end{figure}

\begin{figure}[htbc]
\includegraphics[width=0.75\textwidth]{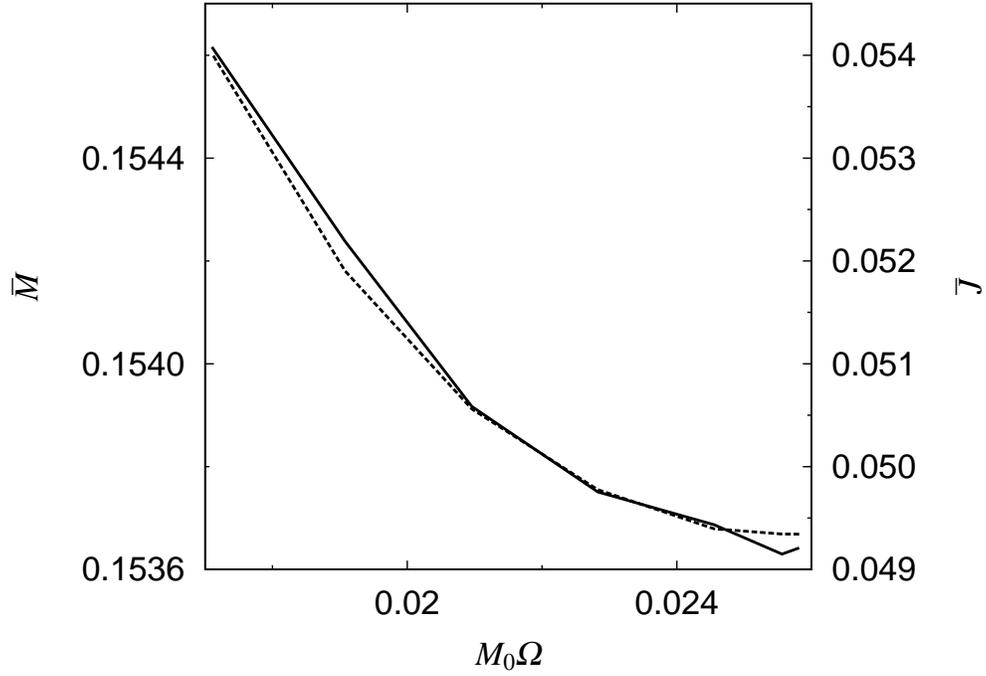}
\caption{Same as Fig.~\protect\ref{dj-dm N=0.5} but for $N=1.0$,
and $(M/R)_{\infty} = 0.175$.
\label{dj-dm N=1.0} 
}
\end{figure}

In Figs.~\ref{dj-dm N=0.5} and \ref{dj-dm N=1.0}, the nondimensional 
gravitational mass and the nondimensional angular momentum of each star 
are plotted against the nondimensional angular velocity. From these figures, 
it can be seen that turning points, i.e., the minima of each value, 
of two curves coincide. It implies that secular stability of binary
systems can be found by investigating either the gravitational mass
or the angular momentum.

This is a nice feature that agrees the requirement between the change of 
the gravitational mass and that of the angular momentum as follows:
\begin{equation}
dM = \Omega dJ \ ,
\label{dj-dm}
\end{equation}
where $dM$ and $dJ$ are the changes of the gravitational mass and the
angular momentum of two configurations with the same rest mass, respectively.
This relation can be reduced from the first law of thermodynamics, which 
is shown below, for the binary systems for which the rest mass, entropy, 
and vorticity of each fluid element are conserved~\cite{FKM01}:
\begin{equation}
dE = \Omega dJ \ ,
\label{dj-de}
\end{equation}
where $E$ means the half of the total energy of the system.

It should be noted that these requirements can be checked if we can 
obtain highly accurate models. As seen from Figs.~\ref{dj-dm N=0.5}
and \ref{dj-dm N=1.0}, changes of the gravitational mass and
the angular momentum are three or four orders of magnitude smaller than
the corresponding quantities. Unfortunately, since we cannot insist that
our values have such high accuracy, we do not show our results here.

\begin{figure}[htbc]
\includegraphics[width=0.75\textwidth]{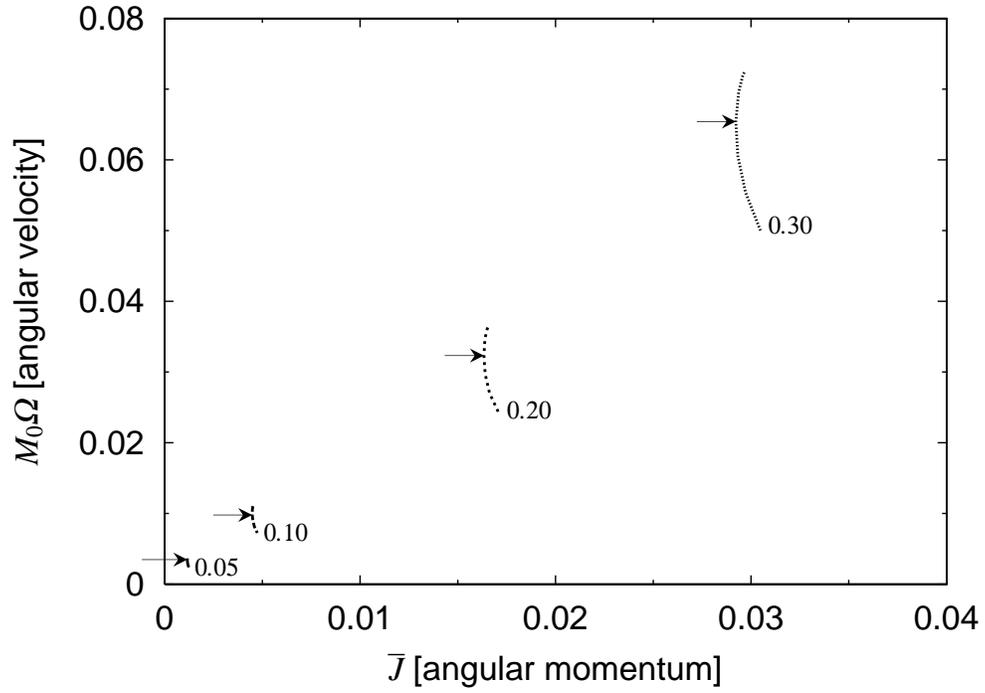}
\caption{Same as Fig.~\protect\ref{J-Omega N=1.0} but for $N=0.5$. 
Four curves correspond to sequences with the compactness
$(M/R)_{\infty} = 0.05$, $0.10$, $0.20$, and $0.30$.
Here arrows denote the turning points where the angular momentum becomes minimum.
\label{J-Omega N=0.5}
}
\end{figure}

As seen from Figs.~\ref{J-Omega N=1.0}, \ref{J-Omega N=1.5} and 
\ref{J-Omega N=0.5}, the ranges of the values of $\bar{M}_0 \bar{\Omega}$ and
$\bar{J}$ are considerably wide for the values of $N$.  Even for the 
sequences with the same $N$, the values of $\bar{M}_0 \bar{\Omega}$ and
$\bar{J}$ range widely.  Thus it is not easy to understand the effects
of the strength of gravity and/or the equation of state.

In order to see the features of the evolutionary sequences at a glance, 
we introduce the following two nondimensional quantities, one of
which can be considered to represent the angular velocity and
the other of which corresponds to the angular momentum:
\begin{eqnarray}
\label{New J}
\hat{j} &\equiv& \frac{J}{J_0} \ , \\
\label{New omega}
\hat{\omega} &\equiv& \frac{\Omega}{\Omega_0} \ ,
\end{eqnarray}
where 
\begin{eqnarray}
\label{normalization J}
J_0    &\equiv& \frac{7}{5} G^{1/2} M^2 \left(\frac{M}{R}\right)^{-1/2}
 = \frac{7}{5} M \frac{GM}{c^2} c \left( \frac{GM}{c^2R} \right)^{-1/2} \ , \\
\label{normalization Omega}
\Omega_0 &\equiv& \frac{1}{2} G^{1/2} M^{-1} \left(\frac{M}{R}\right)^{3/2}
 = \frac{1}{2} c \left(\frac{GM}{c^2} \right)^{-1} \left(\frac{GM}{c^2 R} 
\right)^{3/2} \ .
\end{eqnarray}
Here, $R$ means the radius of the star on the major axis measured in the 
Schwarzschild-like coordinate. Our coordinate system in this paper is
a kind of isotropic one and so $R$ is defined as follows:
\begin{eqnarray}
R &=& r_{\rm AB} \left( 1 + \frac{M}{2r_{\rm AB}}\right) ^ 2 \ , \\
r_{\rm AB}  &=& \frac{r_{\rm B} - r_{\rm A}}{2} \ . 
\end{eqnarray}

The meaning of these quantities, ${\hat j}$ and ${\hat \omega}$, can be 
roughly understood if we consider a 
system in Newtonian gravity which consists of two identical rigid spheres 
of uniform density in a contact phase. For such a system, $\hat{j} = 1$ and 
$\hat{\omega} = 1$.  

Another property of these quantities can be seen from the 
definition of the normalization factors, Eqs.~(\ref{normalization J}) and 
(\ref{normalization Omega}). In these expressions, the differences originating 
from the different strength of gravity are ``renormalized" by introducing the 
term related to the quantity $(M/R)_{\infty}$. Thus we will call $\hat{\omega}$ a 
renormalized angular velocity and $\hat{j}$ a renormalized angular momentum.

\begin{figure}[htbc]
\includegraphics[width=0.75\textwidth]{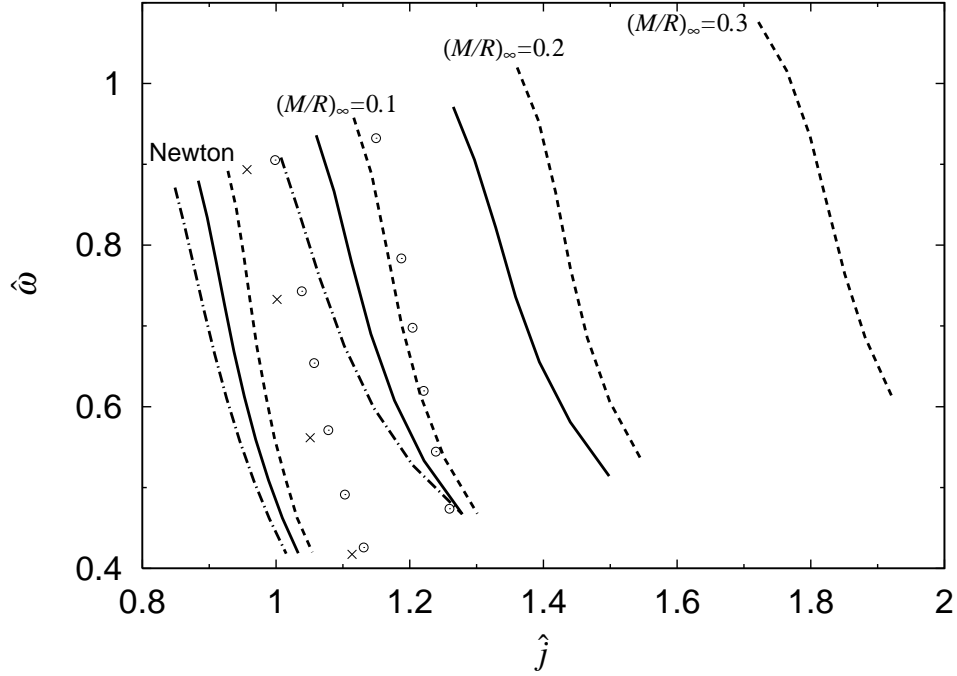}
\caption{Renormalized angular velocity $\hat{\omega}$ 
(defined as Eq.~(\protect\ref{New omega})) is plotted against
renormalized angular momentum $\hat{j}$ 
(defined as Eq.~(\protect\ref{New J}))
for the sequences of constant rest mass binary systems
with different values of the polytropic indices: $N=0.5$ (dashed curves), 
$N=1.0$ (solid curves), and $N=1.5$ (dash-dotted curves). Numbers attached
to each curve are the values of $(M/R)_{\infty}$.
Sequences of Baumgarte {\it et al.}~\protect\cite{BCSST98a} are 
also drawn by different symbols, as
open circles for $N=1.0$ and $(M/R)_{\infty} = 0.10$, and $0.20$, and 
crosses for $N=1.5$ and $(M/R)_{\infty} = 0.10$.
It should be noted that there exist no quasiequilibrium sequences
with $(M/R)_{\infty} = 0.2$ and $0.3$ for $N = 1.5$ polytropes
and with $(M/R)_{\infty} = 0.3$ for $N = 1.0$ polytropes
because even the maximum mass configurations cannot
reach the corresponding values of $(M/R)_{\infty}$ for
$N = 1.0$ and $1.5$ polytropes.
\label{J-Omega New}
}
\end{figure}

In Figs.~\ref{J-Omega New}, the renormalized angular velocity $\hat{\omega}$ 
is plotted against the renormalized angular momentum $\hat{j}$ for 
several sequences of $N$ and $(M/R)_{\infty}$. As seen from this figure, 
the values of $\hat{\omega}$ and $\hat{j}$ for all evolutionary sequences 
with constant rest masses are scaled to values around unity. For the 
Newtonian sequences, the position of contact phases for smaller values of $N$ 
approaches $(1.0, 1.0)$ but never reaches that point because configurations 
are not rigid bodies and deformed from spheres by the tidal force from the 
companion star. 

Several characteristic features can be seen in this figure. First, 
if we compare the sequences with the same value of $(M/R)_{\infty}$, sequences 
with stiffer equations of state locate at the upper-right region. 
This can be explained as follows. If we choose models which have the
same values of the gravitational mass and the angular velocity, 
the radii are the same so that the inertial moment is larger for the stiffer 
polytropes.  It implies that the angular momentum is larger for stiffer 
equations of state.  Concerning the value of the renormalized angular 
velocity at the contact stage, the gravitational force is stronger
for the stiffer polytropes because of the distribution of the matter
inside the star. Thus larger angular velocity is required for configurations 
with stiffer equations of state.

Second, if we compare the sequences with the same value of $N$, 
more relativistic sequences locate at the larger values of $\hat{j}$.
This is explained as follows. If we choose the models which have
the same values of $N$, $M$ and $\hat{\omega}$, we obtain the following
relation:
\begin{equation}
\hat{j} \propto \frac{< (r \sin \theta)^2 >}{R^2} \ ,
\end{equation}
where $<F>$ is the mass average of the quantity $F$ defined as
follows:
\begin{equation}
< F > \equiv  \frac{\int F dm}{\int dm} \ .
\end{equation}
Here $dm$ is the mass element of the configuration.
Since, in general, the change of the averaged quantity is smaller than the
change of the quantity itself, the change of $\hat{j}$ is affected mainly 
by the change of $R$ which decreases as $(M/R)_{\infty}$ increases.  Therefore the 
value of $\hat{j}$ increases and the curves are shifted towards right in the 
plane.

It should be noted that if we consider sequences with the same value
of $(M/R)_{\infty}$ but different values of $N$, differences due to 
different values of $N$ are amplified for configurations with the larger 
value of $(M/R)_{\infty}$.  Since real neutron stars 
cannot be approximated by a single polytropic relation all through the
whole star, the evolutionary sequences cannot be approximated by the 
assumptions adopted in this paper, i.e., the assumption that
$N$ and $K$ are conserved. Therefore, in order to get information
about real evolutions, we need to construct evolutionary sequences 
with realistic equations. \\

We would like to thank Dr. K\=oji Ury\=u for his helpful discussions.
FU is a Research Fellow of the Japan Society for the Promotion of Science 
(JSPS) and is grateful to JSPS for the financial support. This work was 
partially supported by the Grant-in-Aid for Scientific Research (C) of 
JSPS (12640255).



%
%

%
%

%
%

%
%

%
%

%
%

%
%

\end{document}